\documentclass[10pt]{article}
\usepackage{amsfonts}
\usepackage{epsfig}
\usepackage{cite}
\begin{document}
\begin{center}
{\Large\bf Two-photon Exchange Corrections to Single Spin \\
\vspace{0.3cm}
Asymmetry of Neutron and $^3$He}\\%
\vspace*{1cm}
Dian-Yong Chen $^{1,2}
  \footnote{E-mail: chendy@impcas.ac.cn}$,
Yu-Bing Dong$^{3,4}$\\
\vspace{0.3cm} %
{\ $^1$Research Center for Hadron and CSR Physics, Lanzhou
University\\
   $\&$ Institute of Modern Physics of CAS, Lanzhou 73000, China }\\
{\ $^2$Institute of Modern Physics, \\
Chinese Academy of Science,\ Lanzhou,\ 730000, China}\\
{\ $^3$Institute of High Energy Physics, \\
Chinese Academy of Science,\ Beijing,\ 100049, China}\\
{\ $^4$ Theoretical Physics Center for Science Facilities, CAS,
Beijing 100049, China}
\vspace*{1cm}
\end{center}

\begin{abstract}
In a simple hadronic model, the two-photon exchange contributions to
the single spin asymmetries for the nucleon and the $^3$He are
estimated. The results show that the elastic contributions of
two-photon exchange to the the single spin asymmetries for the
nucleon are rather small while those for the $^3$He are relatively
large. Besides the strong angular dependence, the two-photon
contributions to the single spin asymmetry for the $^3$He are very
sensitive to the momentum transfer.
\end{abstract}
\section{Introduction}
In recent years, the two-photon exchange (TPE) effect on the
electron-nucleon scattering process attracts many interests again
after its success in recoiling the discrepancy of the form factor
ratio caused by different experimental extraction techniques
\cite{PRL-142304, PRL-142303, PRL-122301, PRL-172503, PRC-038201,
PRD-013008}. The newly estimate results showed that the TPE
correction is rather important in extracting the proton's form
factors from Rosenbluth separation because of its explicit angle
dependence. In addition to explaining the form factor ratio
discrepancy, further calculations have emphasized the direct
connection between the TPE process and the single spin asymmetry
(SSA), $A_y$. For the elastic scattering process, this asymmetry is
expected to be vanishing in the one photon exchange approximation
due to time-reversal invariance. However, it can receive non-zero
contributions from the interference between the one-photon exchange
amplitude and the imaginary part of the TPE amplitude. That means a
nonzero $A_{y}$ should be a strong evidence for the exist of the TPE
contributions. Since the Born contribution is not present, the
measurements of $A_{y}$ would provide a unique opportunity to access
the information on nucleon structure through the dynamics of the
TPE. Recently, in the Jefferson Lab, an experiment was designed to
measure the single spin asymmetry of neutron using a vertically
polarized $^3$He target\cite{E05-015}. Since free neutron targets do
not exist in nature, one has to consider nucleon bound systems, such
as deuteron \cite{NPA-740} or $^3$He\cite{PRC-538}. For the latter
case, within a naive model and with only a symmetric S-wave
component in the bound state, the two protons have opposite spins
and therefore one should expect that the electromagnetic polarized
response of the $^3$He essentially is the neutron one.
\par%
Theoretically, some calculations about the TPE corrections to the
SSAs for nucleon target have been performed in both the simple
hadronic model \cite{PRC-034612} and the parton model
\cite{PRD-013008}. In the latter approach, the results showed that
at $Q^2=6\ GeV^2$ the TPE contributions to the SSA for the proton is
of order $1\%$, and for the neutron, at $Q^2=4.3\ GeV^2$, the SSA is
somewhat larger than those for the proton \cite{PRD-013008}. In
large momentum transfer region, the electron-nucleon scattering can
be considered through the scattering off partons in the nucleon, and
well described by the generalized parton distribution functions.
However, the experiment designed in the Jefferson Lab was concerned
to the single spin asymmetry for the neutron under $1\ GeV^2$. In
such low $Q^2$ range, we believe the simple hadronic model should be
more efficient to consider the TPE contributions. In this model, the
SSA for the proton target has been evaluated \cite{PRC-034612} and
in present work we further estimate the TPE corrections to the SSAs
for the neutron and the $^3$He targets.
\par%
This paper is organized as follows. We present some analytical
representations of the TPE contributions to the SSAs for the nucleon
and the $^3$He within the simple hadronic model in the following
section. In section \ref{Chap-Num}, some numerical results and
discussions about the TPE contributions to the SSAs for the nucleon
and the $^3$He targets are presented.

\section{TPE Contributions to the SSAs for Neutron and $^3$He}
\label{Chap-TPE}%

The $^3$He is also spin$-1/2$ particle, thus one can discuss the TPE
contributions to the $^3$He in the same way as the one has been done
for the nucleon. Here taking electron-nucleon scattering process
$e^{-}(p_{1}) + N(p_2) \rightarrow e^{-}(p_3) +N(p_4) $ for example,
as one knows, after considering the TPE contributions, an extra term
is introduced in the effective electromagnetic vertex of the nucleon
and the vertex becomes:
\begin{eqnarray}
\Gamma^{\mu}= \widetilde{F}_{1} \gamma^{\mu} +\widetilde{F}_2
\frac{i \sigma^{\mu\nu} q_{\nu}}{2M} + \widetilde{F}_3 \frac{\gamma
\cdot K P^{\mu}}{M^2}
\label{Eq-FFs-2g}%
\end{eqnarray}
with $M$ is the nucleon mass and the three independent momenta are
$q=p_{1}-p_{3}$, $K=p_{1}+p_{3}$ and $P=p_{2}+p_{4}$. The form
factors $\widetilde{F}_{i}, \{i=1,2,3\}$ in above vertex are the
functions of both the square of the momentum transfer $Q^2$ and the
photon polarization parameter $\epsilon$, which is related to the
scattering angle by $\epsilon= [1 + 2(1+\tau) \tan^2
(\theta/2)]^{-1}$. The form factors $\widetilde{F}_{1,2}$ are
usually recombined as $\widetilde{G}_{E,M}$ and $\widetilde{F}_3$ is
expressed by $Y_{2\gamma}$. Moreover, the TPE contributions are
separated from those based on one photon approximation, and one
have,
\begin{eqnarray}
\widetilde{G}_{E,M} (Q^2, \epsilon) = G_{E,M} (Q^2) +\Delta G_{E,M}
(Q^2, \epsilon),\ \ \  Y_{2\gamma} = \frac{\nu}{M^2}
\frac{\widetilde{F}_{3}}{ G_{M}},
\end{eqnarray}
with $\nu= P \cdot K$ and $G_{E,M}$ are the electromagnetic form
factors under one-photon approximation. With the contributions of
the TPE process, the reduced differential cross section and the SSA
can be expressed as:
\begin{eqnarray}
\sigma_{R} &=& G_{M}^2 +\frac{\epsilon}{\tau} G_{E}^2 + 2 G_{M}
\mathcal{R} \Big( \Delta G_M + \epsilon G_{M} Y_{2 \gamma}\Big) + 2
\frac{\epsilon}{\tau} G_{E} \mathcal{R} \Big( \Delta G_{E} +G_{M}
Y_{2 \gamma} \Big) +\mathcal{O}(e^4),\nonumber\\[3pt]
A_{y} &=& \sqrt{\frac{2\epsilon(1+\epsilon)}{\tau}}
\frac{1}{\sigma_{R}} \Big\{ -G_{M} \mathcal{I} \Big( \Delta G_{E}
+G_{M} Y_{2 \gamma}\Big) +G_{E} \mathcal{I} \Big( \Delta G_{M} +
\Big( \frac{2\epsilon}{1+\epsilon}\Big) G_{M} Y_{2 \gamma} \Big)
\Big\}.
\label{Eq-DCS-Ay}%
\end{eqnarray}
\par
\begin{figure}
\centering
\mbox{\epsfig{figure=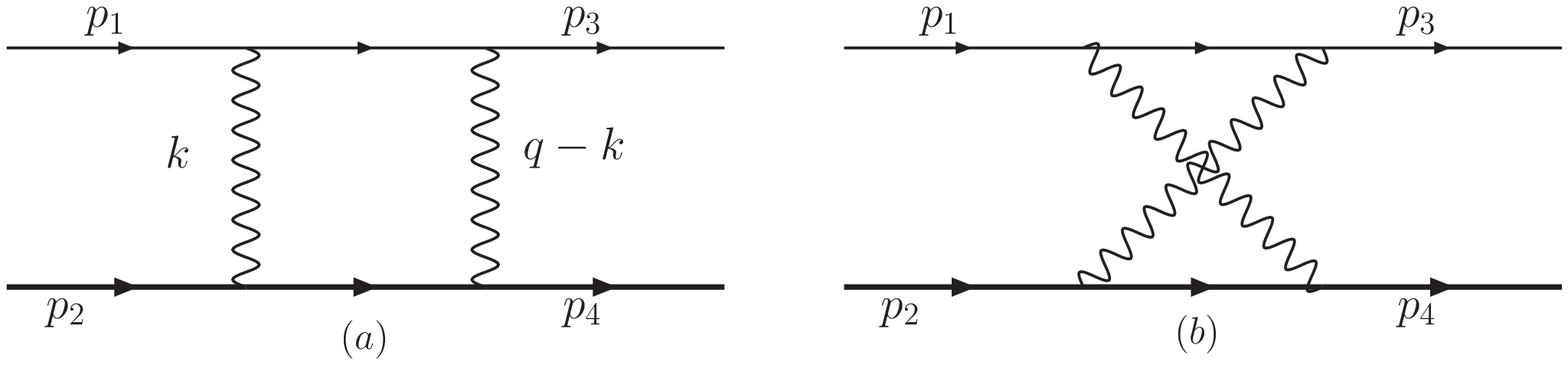,width=140mm,clip=}} %
\renewcommand{\figurename}{Fig.}
\caption{Feynman diagram used in present calculations.}
\label{Fig-Feyn-TPE}%
\end{figure}

%
In this work, we restrict ourself to low energy transfer region with
$Q^2<1\ GeV^2$. Then in the calculations of the TPE corrections we
only include the nucleon as the intermediate state. The Feynman
diagrams are shown in Fig. \ref{Fig-Feyn-TPE}. The TPE amplitudes
corresponding to the diagrams are:
\begin{eqnarray}
\mathcal{M}^{2 \gamma}= Z^2 e^4 \int \frac{d^4 k}{(2 \pi)^2} \Big[
\frac{N_{a}(k)}{D_{a}(k)} +\frac{N_{b}(k)}{D_{b}(k)} \Big].
\label{Eq-Amp-2g}%
\end{eqnarray}
Here $Z$ is the charge number of the targets, for the nucleon $Z=1$
and for the $^3$He $Z=2$. For the box diagram (as shown in Fig.
\ref{Fig-Feyn-TPE} $(a)$), one has,
\begin{eqnarray}
N_{a} &=& \bar{u}(p_3) \gamma_{\mu} (\hat{p}_1- \hat{k})
\gamma_{\nu} u(p_1) \bar{u}(p_4) \Gamma^{\mu}_{1\gamma}(q-k)
(\hat{p}_2 +\hat{k} +M) \Gamma^{\nu}_{1\gamma}(k) u(p_2),\nonumber\\[5pt]
D_{a} &=& [k^2 -\lambda^2] [(k-q)^2-\lambda^2] [(p_1-k)^2-m^2]
[(p_2+k)^2-M^2].
\label{Eq-NaDa}%
\end{eqnarray}
and in the same way one can get the expressions of $N_{b}(k)$ and
$D_{b}(k)$ for the crossed box diagram (as shown in Fig.
\ref{Fig-Feyn-TPE}(b)). The electromagnetic vertex
$\Gamma^{\mu}_{1\gamma}(q)$ employed in Eq. (\ref{Eq-NaDa}) is the
one under one-photon approximation. The form factors $F_{1}$ and
$F_{2}$ are directly parameterized in terms of the sums of
monopoles, of the form \cite{PRC-034612},
\begin{eqnarray}
F_{1,2} (Q^2) = \sum_{i=1}^{N} \frac{n_i}{d_{i}+Q^2},
\label{Eq-FFsFit}%
\end{eqnarray}
where $n_{i}$ and $d_{i}$ are free parameters, and the normalization
conditions are $F_{1}^{p}=1$ and $F_{2}^{p} =\kappa_{p}$ for the
proton, $F_{1}^{n}=0$ and $F_{2}^{n}=\kappa_{n}$ for the neutron and
$F_{1}^{He}=1$ and $F_{2}^{He}=\kappa_{He}$ for the $^3$He, where
$\kappa_{p} = 1.793$, $\kappa_{n} = -1.913$ and $\kappa_{He} =
-4.185$ are the proton, neutron and helium anomalous magnetic
moments respectively. For the nucleon, we use the same parameters
given by Ref. \cite{PRC-034612}. The parameters used in this work
for the $^3$He \cite{Comm} together with those for nucleon are list
in Table \ref{Tab-Par}.
\begin{table}[!h]
\caption{Parameters for the nucleon and the $^3$He form factors
fitted in Eq. (\ref{Eq-FFsFit}). $n_i$ and $d_i$ in units of
$GeV^2$.}
\begin{center}
\begin{tabular}{ccccccc}
\hline\hline%
   & $F_{1}^{p}$  & $F_{2}^{p}$ & $F_{1}^{n}$  & $F_{2}^{n}$
   & $F_{1}^{He}$ & $F_{2}^{He}$\\
\cline{2-7}%
$N$&  $3$  &  $3$  &  $3$  &  $2$  &  $3$  &  $3$\\
\hline%
$n_1$ & $0.38676$ & $1.01650$ & $24.8109$ & $5.37640$ & $ 3.58300$ & $ 3.58307$\\
$n_2$ & $0.53222$ & $-19.0246$& $-99.8420$& $5.37640$ & $ 0.19346$ & $ 0.18396$\\
$n_3$ & $-0.94491$& $18.0371$ & $75.0544$ & $ --    $ & $-7.15263$ & $-7.15242$\\
\hline%
$d_1$ & $3.29899$ & $0.40886$ & $1.98524$ & $0.76533$ & $ 0.25033$ & $ 0.23274$\\
$d_2$ & $0.45614$ & $2.94311$ & $1.72105$ & $0.59289$ & $ 3.60308$ & $ 3.58744$\\
$d_3$ & $3.32682$ & $3.12550$ & $1.64902$ & $ --    $ & $ 0.32599$ & $ 0.29276$\\
\hline\hline
\end{tabular}
\end{center}
\label{Tab-Par}
\end{table}
\par%
To carry out the TPE corrections to the SSAs for the nucleon and the
$^3$He, the loop integrals are first evaluated analytically in terms
of the four-point Passarino-Veltman functions \cite{NPB-151} using
the package FeynCalc \cite{CPC-345}. Then, the Passarino-Veltman
functions are evaluated numerically with LoopTools \cite{CPC-153}.
The numerical results about TPE corrections to the SSAs are
displayed in the following sections.
\section{Numerical Results and Discussions}
\label{Chap-Num}
In order to calculate the TPE corrections to the SSAs, we also have
to consider the TPE corrections to the unpolarized differential
cross sections. Similar to previous literatures \cite{PRL-142304,
PRC-034612, PRC-038201}, in the calculations of the unpolarized
differential cross sections, the IR divergences in the TPE loop
integral are canceled out by the standard MT corrections, which were
introduced by L. M. Mo and Y. S. Tsai \cite{PR-1898,RMP-205} and had
been included in the experiment. As one knows, the soft part of the
TPE corrections, which contains all the information on IR
divergences of the loop integral, can be expressed as
$\mathcal{M}_{2\gamma}^{soft} = \delta^{soft} M^{1\gamma}$. That
means the soft part of the TPE corrections contributes to $F_{1}$
and $F_2$ with same factor $\delta^{soft}$. Then correspondingly,
one can conclude that the soft part of the TPE corrections to the
electromagnetic form factors are,
\begin{eqnarray}
\Delta G_{E}^{soft} / G_{E} \equiv \Delta G_{M}^{soft} / G_{M} =
\delta^{soft}.
\label{Eq-TPE-Soft}%
\end{eqnarray}
As shown in Eq. (\ref{Eq-DCS-Ay}), the SSA is related to the
imaginary part of the TPE corrections, and the IR divergent part of
the TPE corrections only contributes to the electromagnetic form
factors. Then we can rewrite it as,
\begin{eqnarray}
A_{y} &=& \sqrt{\frac{2\epsilon(1+\epsilon)}{\tau}}
\frac{1}{\sigma_{R}} \Big\{ G_{E} G_{M} \big[\mathcal{I}( \Delta
G_{M}) /G_{M} -\mathcal{I} (\Delta G_{E}) /G_{E} \big] +G_{M} \big[
\frac{2 \epsilon}{1+\epsilon} G_{E} -G_{M}\big] \mathcal{I} (Y_{2
\gamma}) \Big\}.
\end{eqnarray}
The second squared bracket only relates to $Y_{2\gamma}$ (or
$\widetilde{F}_3$), which is IR finite. Taking Eq.
(\ref{Eq-TPE-Soft}) into the first squared bracket one can see the
soft part of TPE corrections has no contributions to the SSA, that
means the TPE corrections to the terms in the brace are IR finite
and no more considerations about IR divergence are needed.
\par%
As we mentioned in previous section, the expression used to evaluate
the TPE corrections to the $^3$He are the same as those for the
nucleon. In actual calculations, we only replace the nucleon mass,
nucleon charge number and form factors by the corresponding
quantities of the $^3$He in Eqs. (\ref{Eq-FFs-2g}) -
(\ref{Eq-NaDa}). Detailedly, the mass of the $^3$He is about $3$
times larger than the nucleon mass and the charge of the $^3$He is
$2e$. Moreover the electromagnetic form factors of the $^3$He are
somewhat softer than the nucleon form factors and have zeros at $Q^2
\simeq 0.45\ GeV^2$ and $Q^2 \simeq 0.7\ GeV^2$ for the charge and
magnetic form factors respectively \cite{NPA-596}.
\par
\begin{figure}
\centering
\mbox{\epsfig{figure=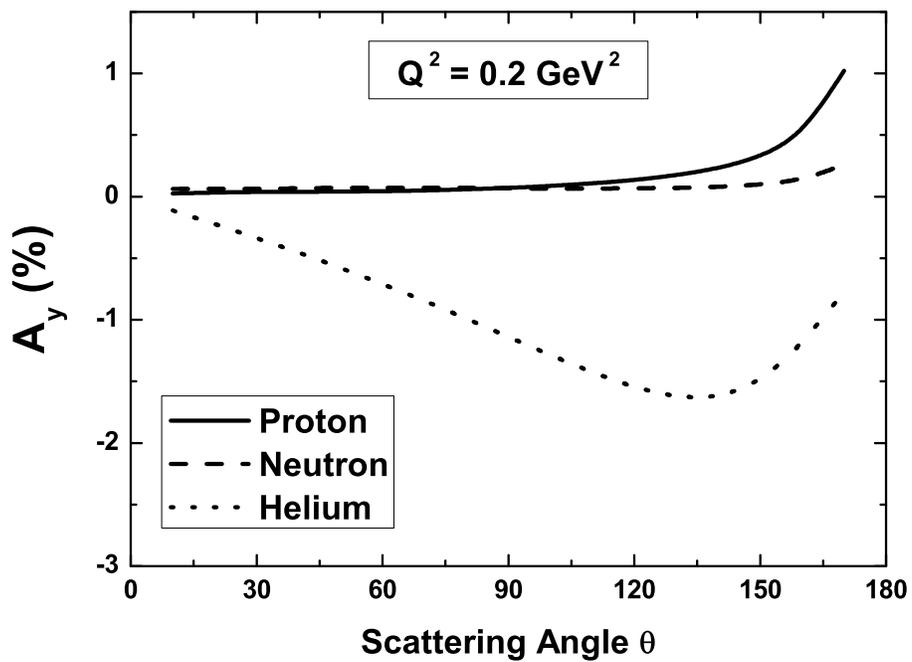,width=120mm,clip=}} %
\renewcommand{\figurename}{Fig.}
\caption{The TPE contributions to the SSAs for the nucleon and the
$^3$He at $Q^2 =0.2\ GeV^2$.}
\label{Fig-Pn-0.2}%
\end{figure}

\begin{figure}
\centering
\mbox{\epsfig{figure=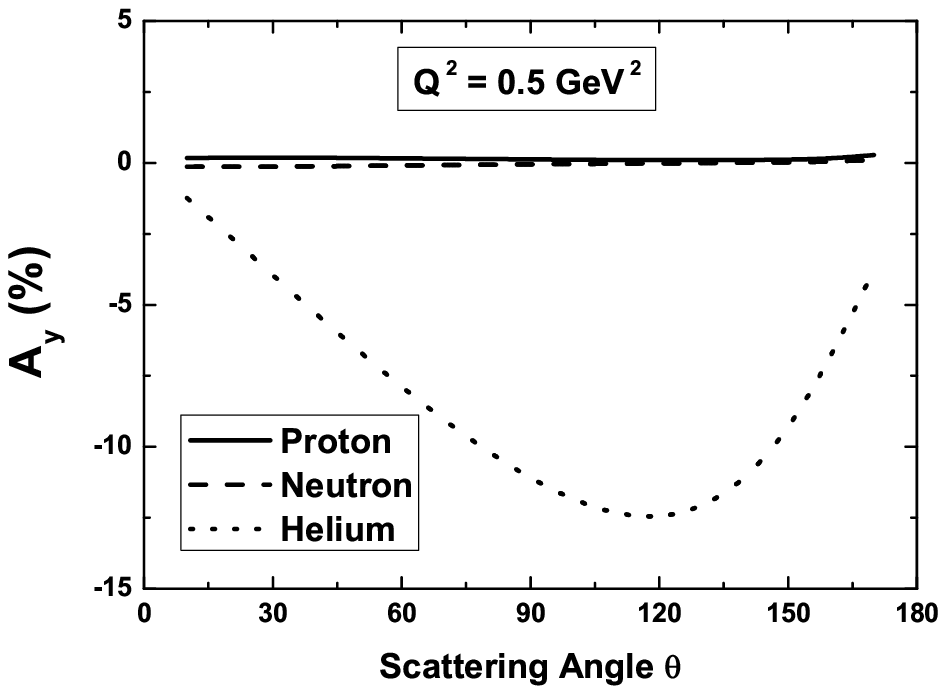,width=120mm,clip=}} %
\renewcommand{\figurename}{Fig.}
\caption{The same as Fig. \ref{Fig-Pn-0.2} but for $Q^2=0.5\
GeV^2$.}
\label{Fig-Pn-0.5}%
\end{figure}

\begin{figure}
\centering
\mbox{\epsfig{figure=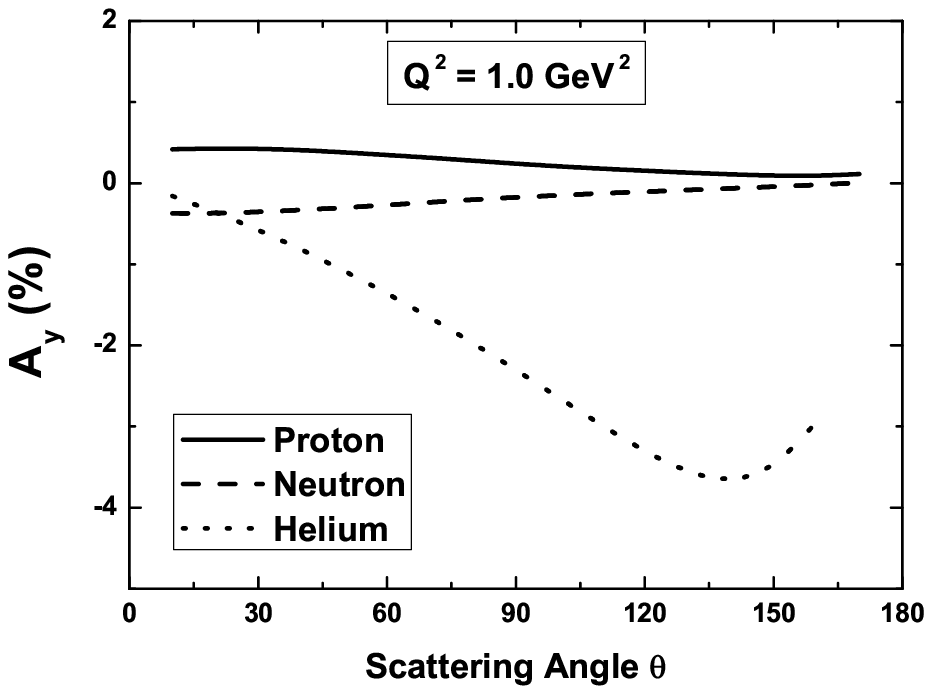,width=120mm,clip=}} %
\renewcommand{\figurename}{Fig.}
\caption{The same as Fig. \ref{Fig-Pn-0.2} but for $Q^2=1.0\
GeV^2$.}
\label{Fig-Pn-1.0}%
\end{figure}

%
The $\theta$ dependence of the TPE contributions to the SSAs for the
nucleon and the $^3$He are displayed in Figs. \ref{Fig-Pn-0.2} -
\ref{Fig-Pn-1.0}. For $Q^2= 0.2\ GeV^2$, as shown in Fig.
\ref{Fig-Pn-0.2}, the TPE contributions to the SSAs for both the
proton and the neutron are positive and increase with the increasing
of the scattering angle $\theta$, and reach about $1\%$ for the
proton and $0.25 \%$ for the neutron in the backward limit. While
the SSA for the $^3$He is negative and decreases with the increasing
of $\theta$ and at about $\theta = 135^\circ$, it reaches minimum
value of approximately $1.6 \%$. For $Q^2= 0.5\ GeV^2$ (Fig.
\ref{Fig-Pn-0.5}) and $Q^2= 1.0\ GeV^2$ (Fig. \ref{Fig-Pn-1.0}), the
TPE contributions to the SSAs for the proton are positive while for
the neutron are negative and moreover, the absolute values decrease
with the increasing of the scattering angle. At the forward limit,
the absolute values of the TPE corrections to the SSAs for the
proton (neutron) are less than $0.2\%$ ($0.15 \%$) at $Q^2=0.5\
GeV^2$ and more than $0.4 \%$ (about $0.4\%$) at $Q^2=1.0\ GeV^2$.
For the $^3$He, the tendency of the curves are similar to the one at
$Q^2=0.2\ GeV^2$ and the minimum values are $15 \%$ and $3 \%$ at
$Q^2=0.5\ GeV^2$ and $Q^2= 1.0\ GeV^2$ respectively. From these
figures, one can conclude the TPE contributions to SSAs for the
$^3$He are strongly dependent on the scattering angle and their
amplitudes are much larger than those for the nucleon. One reason is
that the charge of the $^3$He is $2e$ this produces a factor $4$ in
the TPE amplitudes. Another is that the form factors of the $^3$He
are much softer than those of the nucleon, thus the contributions
from the TPE process with the two photon sharing the momentum
transfer are large.
\par%
In the deep inelastic scattering language, the nucleon structures
are described by four independent functions $W_{1}$, $W_{2}$,
$G_{1}$ and $G_{2}$. The first two are unpolarized structure
functions while the rest two are the polarized ones. Generally, the
polarized structure functions are represented by the dimensionless
functions $g_{1} =M^2 \nu G_{1}$ and $g_{2}= M\nu^2 G_{2}$. In the
deep inelastic region, the structure functions $g_{1,2}$ are only
depend on the Bjorken variable $x= Q^2/2M \nu$. However, this
scaling behavior breaks down in non-perturbative QCD, thus $g_{1,2}$
are functions of both $x$ and $Q^2$. In the naive quark model, where
the quarks are completely independent and gluons are not considered,
$g_{2}$ is expected to be zero. In this case, the SSA, $A_y$, is
directly proportional to the structure function $g_{1}$.
\par%
In general, the spin structure functions of the $^3$He,
$g_{1}^{^3\mathrm{He}}$, can be represented as the convolution of
those of the neutron $g_{1}^n$ and of  the proton $g_{1}^{p}$ with
the spin-dependent nucleon light-cone momentum distributions $\Delta
f_{N/^3\mathrm{He}} (y)$, where $y$ is the ratio of the struck
nucleon to the nucleus light-cone plus components of the momenta
\cite{PRC-968, PRC-38, PRC-024004, PRC-064317},
\begin{eqnarray}
g_{1}^{^3\mathrm{He}} (x,Q^2) =\int_{x}^{3} \frac{dy}{y}\Big[ \Delta
f_{n/^3 \mathrm{He}} (y) g_{1}^{n} (x/y,Q^2) + \Delta f_{p/^3
\mathrm{He}} (y) g_{1}^{} (x/y,Q^2) \Big].
\label{Eq-g1He}%
\end{eqnarray}
Detailed calculations \cite{PRC-968, PRC-38, PRC-024004} by various
groups using different ground-state wave functions of the $^3$He
came to a similar conclusion that $\Delta f_{N/^3\mathrm{He}}$ are
sharply peaked around $y \simeq 1$ due to the small average
separation energy per nucleon. Thus Eq. (\ref{Eq-g1He}) can be
approximated by \cite{PRC-064317},
\begin{eqnarray}
g_{1}^{^3\mathrm{He}} (x, Q^2) =P_{n} g_{1}^{n}(x,Q^2) +2 P_{p}
g_{1}^{p}(x,Q^2),
\label{Eq-g1He-g1N}%
\end{eqnarray}
with $P_{n}$ ($P_{p}$) are the effective polarizations of the
neutron (proton) inside the polarized $^3$He. Considering the ground
state wave function of the $^3$He, which corresponds to $S$-wave
type interaction between any pair of the nucleons inside the $^3$He,
only the neutron is polarized. Thus, one has $P_{n}=1$ and
$P_{p}=0$. In practice, the wave function of the $^3$He include also
higher partial waves, namely the $D$ and $S^{\prime}$ partial waves,
this leads to the depolarization of spin of the neutron and
polarization of  the protons in the $^3$He. The average of
calculations with several models can be summarized as $P_{n}=0.86
\pm 0.02$ and $P_{p} = -0.028 \pm 0.004$ \cite{PRC-2310}.
\par%
According to Eq. (\ref{Eq-g1He-g1N}) and the approximation that the
SSA $A_{y}$ is proportional to the polarized structure $g_{1}$, one
has,
\begin{eqnarray}
A_{y}^{^3\mathrm{He}} =\frac{\sigma^{n}}{\sigma^{^3\mathrm{He}}}
P_{n} A_{y}^{n} + 2 \frac{\sigma^{p}}{\sigma^{^3\mathrm{He}}} P_{p}
A_{y}^{p},
\label{Eq-Ay-He}%
\end{eqnarray}
where $\sigma^{n}$, $\sigma^{p}$ and $\sigma^{^3 \mathrm{He}}$ are
the unpolarized differential cross sections for the neutron, proton
and $^3$He. In present work, we first estimate the TPE corrections
to the SSAs for the $^3$He directly from Eq. (\ref{Eq-Amp-2g}) and
also we can calculate $A_{y}^{^3 \mathrm{He}}$ from Eq.
(\ref{Eq-Ay-He}) based on the TPE corrections to the SSAs for the
nucleon.
\par
\begin{figure}
\centering
\mbox{\epsfig{figure=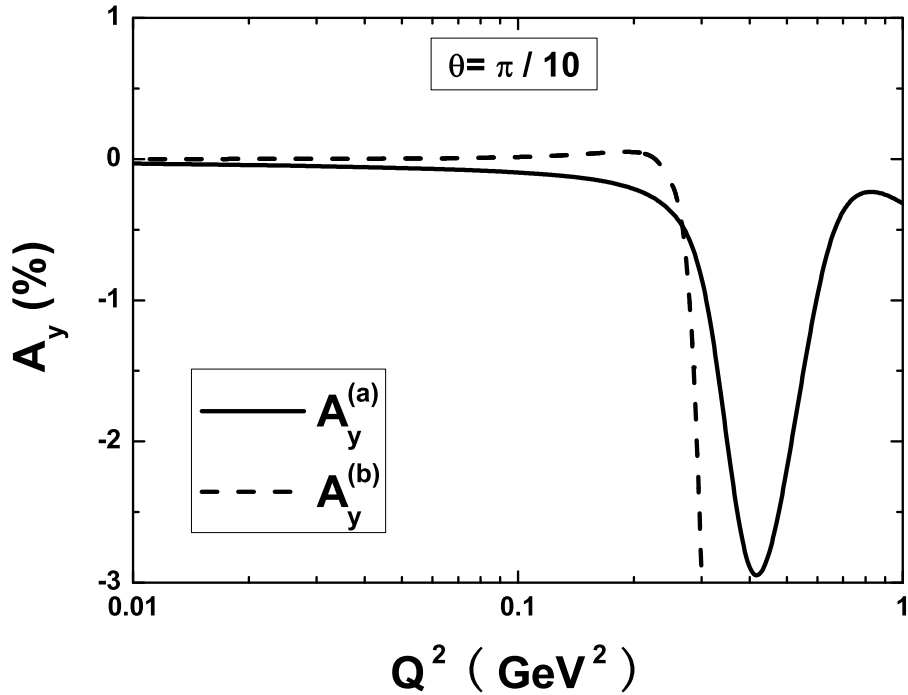,width=120mm,clip=}} %
\renewcommand{\figurename}{Fig.}
\caption{The $Q^2$ dependence of the TPE contributions to the SSA
for the $^3$He at $\theta=\pi/10$ . $A_{y}^{(a)}$ is the result from
a direct estimation of the TPE contributions. while $A_{y}^{(b)}$ is
the result obtained form  Eq. (\ref{Eq-Ay-He}).}
\label{Fig-Pn-Pi-10}%
\end{figure}

\begin{figure}
\centering
\mbox{\epsfig{figure=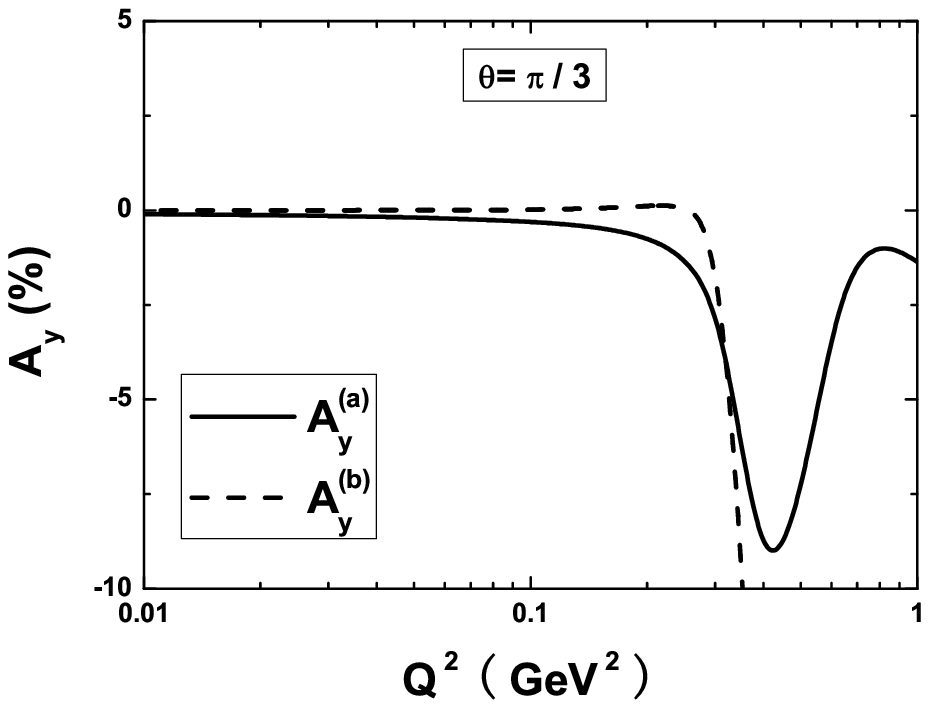,width=120mm,clip=}} %
\renewcommand{\figurename}{Fig.}
\caption{The same as Fig. \ref{Fig-Pn-Pi-10} but for $\theta=\pi
/3$.}
\label{Fig-Pn-Pi-3}%
\end{figure}

\begin{figure}
\centering
\mbox{\epsfig{figure=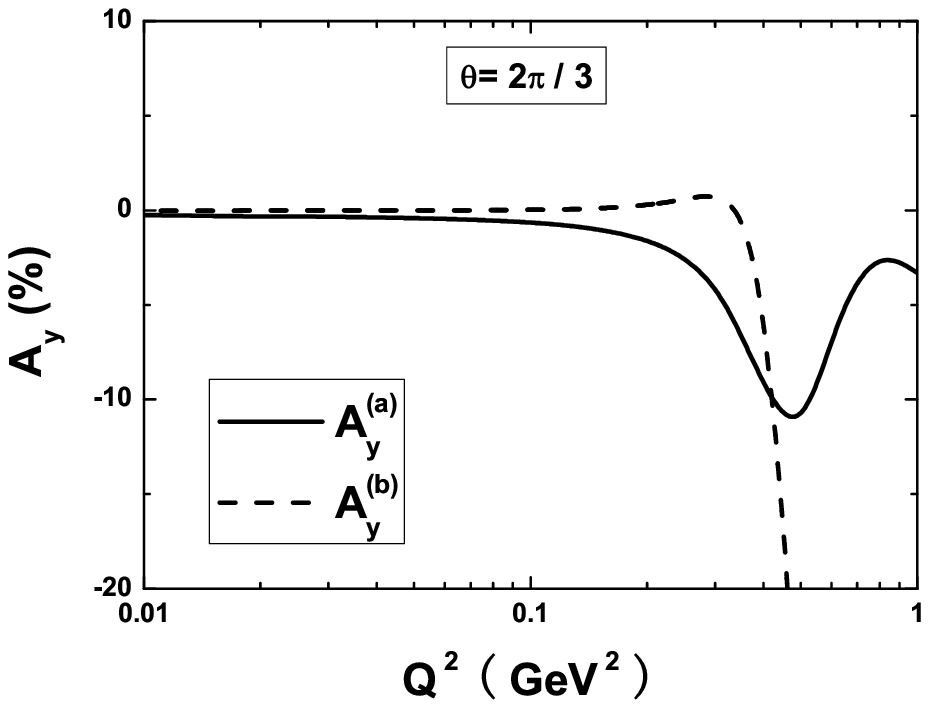,width=120mm,clip=}} %
\renewcommand{\figurename}{Fig.}
\caption{The same as Fig. \ref{Fig-Pn-Pi-10} but for $\theta=2\pi
/3$.}
\label{Fig-Pn-2Pi-3}%
\end{figure}

%
The $Q^2$ dependences of the TPE corrections to the SSAs for the
$^3$He at different scattering angles are displayed in Figs.
\ref{Fig-Pn-Pi-10} - \ref{Fig-Pn-2Pi-3}. In these figures,
$A_{y}^{(a)}$ means the TPE corrections to the SSAs for the $^3$He
obtained directly from Eq. (\ref{Eq-Amp-2g}), which are
corresponding to the solid curves in the figures.  While
$A_{y}^{(b)}$ refers to the SSAs for the $^3$He estimated from Eq.
(\ref{Eq-Ay-He}) based on the TPE corrections to the SSAs for the
nucleon. From these figures one can see, in the region with $Q^2$
under $0.1\ GeV^2$, the SSAs for the $^3$He are very close to zero
and in this region the results obtained from two different methods
are consistent with each other. For the solid curves, one can see,
they all have deep dips in the range $0.4\ GeV^2<Q^2 < 0.5\ GeV^2$.
For $\theta= \pi/10$, the dip appears at $Q^2 \simeq 0.4\ GeV^2$ and
the minimum value is about $3\%$. While for $\theta =\pi/3$, at
$Q^2\simeq 0.43\ GeV^2$, the SSA $A_{y}^{(a)}$ reaches minimum value
of approximately $-8 \%$ and for $\theta = 2\pi/3$, $A_{y}^{(a)}$
reaches its minimum at $Q^2 \simeq 0.5\ GeV^2$ with the value about
$-11\%$. These dips can be well explained by the zero point of the
charge form factor of the $^3$He at $Q^2 \simeq 0.45\ GeV^2$. With
$Q^2$ increasing, the solid curves go up quickly, and in the region
around $0.8\ GeV^2$, they begin to drop again, this can be
interpreted as the reflection of zero point of the $^3$He magnetic
form factor at $Q^2 \simeq 0.7\ GeV^2$.
\par
\begin{figure}
\centering
\mbox{\epsfig{figure=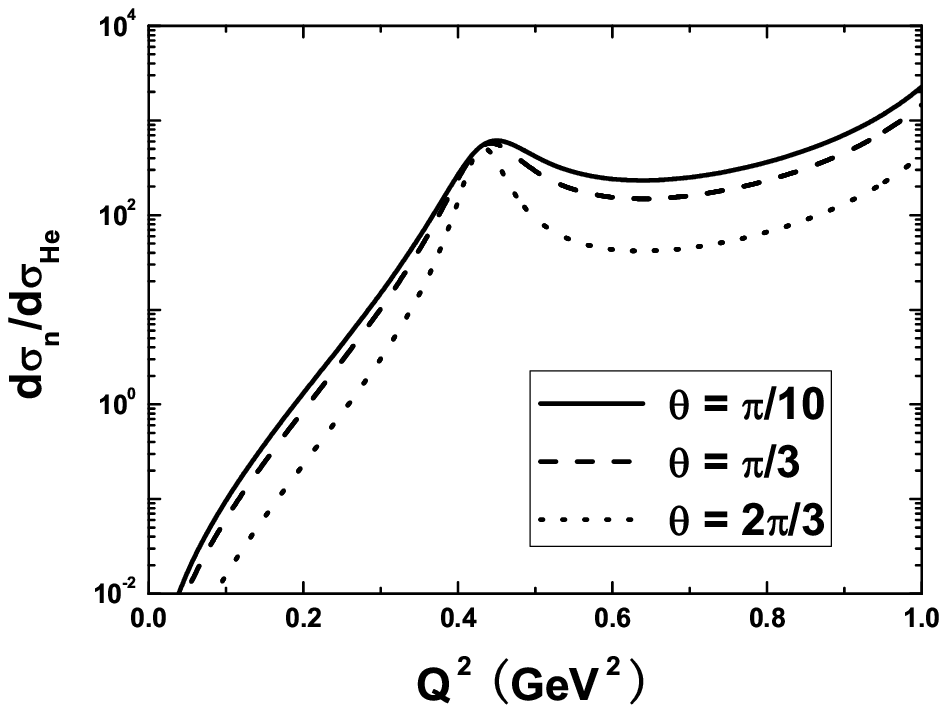,width=120mm,clip=}} %
\renewcommand{\figurename}{Fig.}
\caption{The differential cross section ratio of the neutron and the
$^3$He at the different scattering angle.}
\label{Fig-dcs}%
\end{figure}

%
In Figs. \ref{Fig-Pn-Pi-10} - \ref{Fig-Pn-2Pi-3}, the dashed curves,
which correspond with the SSAs for the $^3$He obtained from Eq.
(\ref{Eq-Ay-He}), drop quickly in the range $Q^2>0.2\ GeV^2$. This
feature is the same with the one of the solid curves. However, as
one can see from these figures, the $A_{y}^{(b)}$ monotonically
decreases with $Q^2$ increasing and quantitatively reaches several
hundred percent at $Q^2= 1\ GeV^2$, which is not consistent with the
solid curves. In Eq. (\ref{Eq-Ay-He}), the effective polarizations
of the proton is only $1/30$ of the neutron, that means, the
polarization of neutron is more important inside the polarized
$^3$He. Thus, one has $A_{y}^{^3\mathrm{He}} \sim \sigma^{n}/
\sigma_{^3\mathrm{He}} P_{n} A_{y}^{n}$. In Fig. \ref{Fig-dcs}, we
shown the differential cross section ratio of the neutron to the
$^3$He at different scattering angles. We see that the form factors
of the $^3$He is much softer than the nucleon (except the neutron
electric form factor), that leads the ratio
$\sigma^{n}/\sigma_{^3\mathrm{He}}$ increases rapidly with the $Q^2$
increasing. From the figure, one can find, in the range $Q^2>0.4\
GeV^2$ the differential cross section of neutron is at least two
orders larger than those of the $^3$He for $\theta =\pi/10$ and
$\pi/3$. Even for $\theta=2\pi/3$, the differential cross section
ratio of neutron to $^3$He is also at least $30$. Such a large ratio
leads to the SSAs for $^3$He obtained from Eq. (\ref{Eq-Ay-He})
decreases rapidly with the increasing of $Q^2$. Above all, in Eq.
(\ref{Eq-Ay-He}), the SSA for the $^3$He is just approximately equal
the the sum of the contributions from the proton and neutron. Eq.
\ref{Eq-Ay-He} had only been verified in the $Q^2$ range from
several $GeV^2$ to about $10\ GeV^2$ \cite{PRC-064317}. The validity
of this approximation keeps unknown in the low energy transfer
region. In Figs. \ref{Fig-Pn-Pi-10}-\ref{Fig-Pn-2Pi-3}, we believe
the results from two different methods are consistent under $0.1\
GeV^2$ just because the TPE contributions are nearly invisible in
such low energy transfer region. Moreover, in the right-hand side of
Eq. (\ref{Eq-Ay-He}), for the calculations of the SSAs for the
nucleon contributed from the TPE process, only the elastic
contribution has been considered. The corrections from the nucleon
resonances, such as $\Delta(1232)$ are expected to revise the
results within a specific limits. Thus, in the present work, we
prefer the results from the direct calculation of Eq.
(\ref{Eq-Amp-2g}). Furthermore, the calculations within the plain
wave impulse approximation \cite{PRC-64, PRC-1591, PRC-38}, in which
the nuclear current arise from the one-nucleon current and the
residual nucleus is assumed not participate in the scattering
precess, may provide more accurate information on TPE corrections to
SSA for $^3$He.
\par%
To summarize, we have studied the TPE corrections to the SSAs for
the nucleon and the $^3$He in a simple hadronic model, and in
present calculations only the elastic contributions are considered.
The SSAs for the nucleon are found to be rather small, while for the
$^3$He are rather large. Another important feature of the SSA for
$^3$He is its strong dependence of the momentum transfer. From our
calculations, one can conclude that the SSAs for the $^3$He should
be more easily measured in the scattering angle range $120^\circ <
\theta < 150^\circ$ and momentum transfer range from $0.3\ GeV^2$ to
$0.6\ GeV^2$ in the experiments. We also try to estimate the SSA for
the $^3$He from the contributions of its nucleon component based on
a very simple approximate equation. The results obtained from this
approximation are comparable to those from the direct calculations
of the TPE process at very small $Q^2$ region. However, large
discrepancies between the results from the two methods are presented
at the range $Q^2> 0.3\ GeV^2$. Further study of the SSAs for the
$^3$He based on the plane wave impulse approximation are needed.

\section{Acknowledgments}
We thank P. G. Blunden and J. P. Chen for helpful discussions and W.
Melnitchouk for his $^3$He form factors parameterizations. This work
is partly supported by the National Sciences Foundations of China
under grants No. 10775148,10975146. The authors are also grateful to
the CAS for grant No KJCX3-SYW-N2.



\clearpage\newpage

\begin{thebibliography}{00}
\bibitem{PRL-142304} P. G. Blunden, W. Melnitchouk and J. A. Tjon,
Phys. Rev. Lett. {\bf 91}, 142304 (2003).
\bibitem{PRL-142303} P. A. M. Guichon and M. Vanderhaeghen, Phys.
Rev. Lett. {\bf 91}, 142303 (2003).
\bibitem{PRL-122301} Y. -C. Chen, A. Afanasev, S. J. Brodsky, C. E.
Carlson and M. Vanderhaeghen, Phys. Rev. Lett. {\bf 93}, 122301
(2004).
\bibitem{PRL-172503} S. Kondratyuk, P. G. Blunden, W. Melnitchouk
and J. A. Tjon, Phys. Rev. Lett. {\bf 95}, 172503 (2005).
\bibitem{PRC-038201} S. Kondratyuk and P. G. Blunden, Phys. Rev. C
{\bf 75},  038201 (2007).
\bibitem{PRD-013008} A. V. Afanasev, S. J. Brodsky, {\it et al}.,
Phys. Rev. D {\bf 72}, 013008 (2005).
\bibitem{E05-015} http://hallaweb.jlab.org/experiment/E05-102/e05-015/index.html
\bibitem{NPA-740} S. Platchkov {\it et al.}, Nucl. Phys. A {\bf 510},
740 (1990).
\bibitem{PRC-538} B. Blankleider and R. M. Woloshyn, Phys. Rev. C {\bf
29} 538 (1984).
\bibitem{PRC-034612} P. G. Blunden, W. Melnitchouk and J. A. Tjon,
Phys. Rev. C {\bf 72}, 034612 (2005).
\bibitem{Comm} W. Melnitchouk and P. G. Blunden, private
communication.
\bibitem{NPB-151} G. Passarino and M. Veltman, Nucl. Phys. B {\bf
160}, 151 (1979).
\bibitem{CPC-345} R. Mertig, M. Bohm and A. Denner, Comput. Phys.
Commun. {\bf 64}, 345 (1991).
\bibitem{CPC-153} T. Hahn, M. Perez-Victoria, Comput. Phys.Commun.
{\bf 118}, 153 (1999).
\bibitem{PR-1898} Y. S. Tsai, Phys. Rev. 122 (1961) 1898.
\bibitem{RMP-205} L. W. Mo and Y. S. Tsai, Rev. Mod. Phys. 41
(1969) 205.
\bibitem{NPA-596} A, Amroun {it et al}., Nucl. Phys. A {\bf 579},
596 (1994).
\bibitem{PRC-968} C. Ciofi degli Atti, S. Scopetta, E. Pacce, and G.
Salme, Phys. Rev. C {\bf 48}, 968 (1993).
\bibitem{PRC-38} R. W. Sch\"{u}ltze and P. U. Sauer, Phys. Rev.
C {\bf 48}, 38 (1993).
\bibitem{PRC-024004} F. Bissey, A. W. Thomas, and I. R. Afnan, Phys.
Rev. C {\bf 64}, 024004 (2001).
\bibitem{PRC-064317} F. Bissey, V. Guzey, M. Strikman, and A.
Thomas, Phys. Rev. C {\bf 65}, 064317 (2002).
\bibitem{PRC-2310} J. L. Friar, B. F. Gibson, G. L. Payne, A. M.
Bernstein, and T. E. Chupp, Phys. Rev. C {\bf 42} 2310 (1990).
\bibitem{PRC-64} A. Kievsky, E. Pace, G. Salme, and M. Viviani,
Phys. Rev. C {\bf 56}, 64 (1997).
\bibitem{PRC-1591} C. Ciofi degli Atti, E. Pace, and G. Salme, Phys.
Rev. C {\bf 46}, R1591 (1992); {\bf 51}, 1108 (1995)
\end{thebibliography}
\end{document}